\documentclass[%
 reprint,
superscriptaddress,
 amsmath,amssymb,
 aps,
 pra,
]{revtex4-2}

\usepackage{graphicx}
\usepackage{dcolumn}
\usepackage{bm}
\usepackage{stmaryrd}
\usepackage{amsmath}
\usepackage{amssymb}
\usepackage{textcomp}
\usepackage{calrsfs}
\usepackage{yfonts}
\usepackage{amsthm}
\usepackage{lipsum}
\usepackage{color}
\usepackage{subfigure}
\usepackage[section]{placeins}

\begin{document}
\preprint{APS/123-QED}

\title{Electromagnetic field quantization in the presence of a moving nanoparticle}
\author{Vahid Ameri}
\email{vahameri@gmail.com }
\affiliation{Department of Physics, Faculty of Science, University of Hormozgan, Bandar-Abbas, Iran}
\author{Alidad Askari}
\affiliation{Department of Physics, Faculty of Science, University of Hormozgan, Bandar-Abbas, Iran}
\author{Morteza Rafiee}
\affiliation{Faculty of Physics, Shahrood University of Technology, P.O. Box 3619995161 Shahrood, Iran}
\author{Mohammad Eghbali-Arani}
\affiliation{Department of Physics, University of Kashan, Kashan, Iran}

\date{\today}

\begin{abstract}
An appropriate Lagrangian is considered for a system comprising a moving nanoparticle in a semi-infinite space, and the electromagnetic and matter fields are quantized. Through an analysis of the absorbed power radiation, it is demonstrated that the quantum friction experienced by high-velocity nanoparticles can be identified as a dissipative term in the radiation power of the nanoparticle. The absorbed power radiation for a moving nanoparticle is derived and compared with that of a static one. By considering two different temperature scenarios, it is explicitly shown that the absorbed power radiation for a moving nanoparticle always contains a negative term in its power spectrum, which can be attributed to the power lost due to non-contact quantum friction.
\end{abstract}

\maketitle


\section{Introduction}
Ignoring the influence of the surrounding environment in the study of physical systems, whether in a quantum or classical state, can lead to results that significantly differ from those obtained in laboratory settings. This is particularly relevant in a broad range of physical systems where interactions with the electromagnetic field and matter field of the environment are crucial. For example, this can be observed in open quantum systems. Failure to consider the impact of the environment in these systems can result in outcomes that deviate significantly from those observed in laboratory experiments.\cite{PhysRevA.11.230,breuer2002theory,PhysRevA.105.032406}. 
\par 
The process of quantizing the electromagnetic field in the presence of matter fields is another instance of examining the impact of environmental fluctuations on physical systems. Moving objects will also encounter intriguing quantum effects as a result of their motion in the electromagnetic vacuum field. The renowned Casimir force, which arises from the interaction of electromagnetic and matter field fluctuations, is a direct consequence of this phenomenon.\cite{ameri2016perturbative,Dalvit2011,bordag2009advances}. In recent studies, the significance of the Casimir effect in nanoscale devices, particularly in optoelectronic and optomechanical systems, has garnered significant attention. \cite{velichko2020casimir,nie2014generating,PhysRevX.8.011031}. Furthermore, Studies have demonstrated that the rotation of nanoparticles (NPs) can affect both the rate and spectrum of their heat transfer \cite{ameri2017effect,ameri2015radiative,dedkov2016radiation}.  
The force of quantum friction, which arises between two stationary objects that are not physically touching but are in relative motion, is a result of the interaction between the fluctuations in the matter field. \cite{dedkov2020quantum,Dalvit20112,dedkov2020nonlocal}. In fact, there are studies that have shown the presence of a frictional torque on rotating NPs. even in an electromagnetic vacuum, which is not due to any external force. \cite{PhysRevA.82.063827,PhysRevLett.105.113601,Ameri:17,PhysRevA.89.032124}. 
\par 
The non-contact friction force of moving nanoparticles has been investigated in a various settings.
\cite{silveirinha2014theory, philbin2010canonical,horsley2012canonical,zhao2012rotational}.
The objective of this study is to examine a system where non-contact friction arises from the interaction with the electromagnetic vacuum field in the presence of an infinite, ideal conductor. Unlike conventional approaches to investigate quantum friction, which involve two bodies in relative motion and include interaction terms between the bodies and the electromagnetic vacuum and between the bodies themselves, this work only considers the interaction between the NP object and the electromagnetic vacuum, with no other interaction terms. The role of the semi-infinite bulk is to serve as an ideal conductor, with a minimal contribution, and simply reflect the electromagnetic field, represented by a Green function.

 In this study, we explore the behavior of electromagnetic and matter fields surrounding a NP as it moves parallel to the surface of a semi-infinite bulk. Our approach involves quantizing these fields, which allows us to accurately predict the absorbed radiation power of the NP and the dissipation caused by quantum friction between the NP and the semi-infinite bulk. This approach allows us to better understand the behavior of these fields and their interactions with matter.\\

By quantizing the fields surrounding the NP, we are able to calculate the amount of radiation that is absorbed by the particle as it moves through the bulk. This absorbed radiation power is a crucial factor in determining the motion of the NP, as it affects the force acting on the particle.\\

In addition to calculating the absorbed radiation power, we also determine the dissipation caused by quantum friction between the NP and the bulk. This dissipation arises due to the interaction between the quantized fields and the matter surrounding the NP, and can significantly impact the motion of the particle.\\

Overall, our work provides a more detailed understanding of the behavior of NPs in their environment, which is crucial for a variety of applications in fields such as nanotechnology, biotechnology, and materials science. By quantizing the electromagnetic and matter fields surrounding these particles, we are able to more accurately predict their motion and behavior, which can lead to the development of more advanced and effective technologies.

\section{\label{sec:level2}Lagrangian}
\begin{figure}[h!]
\begin{center}
\includegraphics[scale=0.5]{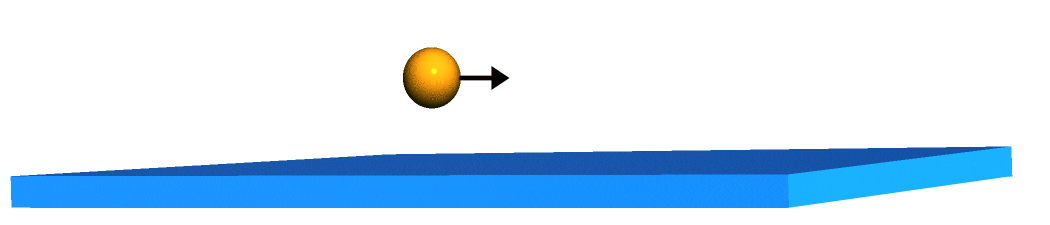}\\
\end{center} 
\caption{Moving nanoparticle (with the radius a=10 nm) parallel to the surface of a semi-infinite bulk in the presence of electromagnetic vacuum field. }\label{setup}
\end{figure}

Let us refer to Fig.(\ref{setup}) for the description of our system, which consists of NP with an initial velocity of $V_0$ moving along the $x$-axis and an semi-infinite bulk. The coordinate-derivative transformations that relate moving and stationary coordinates can be expressed as
\begin{eqnarray}\label{T}
&& z=z',\,\,\,y=y',\,\,\,x=\gamma (x' + V_0t'),\,\,\,t=\gamma(t'+\frac{V_0x'}{c^2}),\nonumber\\
&& \partial_{z'}=\partial_{z},\,\,\,\partial_{y'}=\partial_{y},\,\,\,\partial_{x'}=\gamma(\partial_{x}+\frac{V_0}{c^2}\partial_{t}),\,\,\,\nonumber\\
&&\partial_{t'}=\gamma(\partial_{t}+V_0\partial_{x}),  
\end{eqnarray}
where $c$ is the speed of light and $\gamma = 1/\sqrt{1-v^2/c^2}$ is the Lorentz factor, we can introduce a Lagrangian for the electromagnetic field and matter field of a moving NP in a moving frame, where the prime over space and time notation denotes coordinates in this moving frame. This approach allows us to consider the behavior of these fields in the context of special relativity.
\begin{eqnarray}\label{L}
\mathcal{L} =&& \frac{1}{2}\epsilon_0\,(\partial_t \mathbf{A})^2-\frac{1}{2\mu_0}(\nabla\times\mathbf{A})^2\nonumber\\
&&+\frac{1}{2}\int_{0}^{\infty}d\nu\,[\gamma^2(\partial_t\mathbf{X}+V_0\partial_{x}\mathbf{X})^2-\nu^2\mathbf{X}^2]\nonumber\\
&&- \epsilon_0\int_{0}^{\infty} d\nu\,f_{ij}(\nu,t)X^j\partial_t A_i\nonumber\\
&&+\epsilon_0\int_{0}^{\infty} d\nu\,f_{ij}(\nu,t)X^j (\mathbf{v}\times\nabla\times\mathbf{A})_i.   
\end{eqnarray}
The first two terms of the Lagrangian correspond to the electromagnetic field, while the third term is the modified Lagrangian of the moving NP object due to the transformations given by Eq. (\ref{T}). The last two terms represent the interactions between the matter field of the NP and the electromagnetic field, where $ X^j $ denotes the matter field of the NP and the coupling tensor $ f_{ij} $ is defined as follows:
\begin{equation}\label{C}
f_{ij} (\nu,t)=\left(
\begin{array}{ccc}
    f_{xx} (\nu) & 0 & 0 \\
    0 & f_{yy} (\nu) & 0 \\
    0 & 0 & f_{zz}(\nu)\\
  \end{array}
\right).
\end{equation}
\section{\label{sec:level3}Quantization and equations of motion}

From Lagrangian Eq.(\ref{L}) we find the corresponding conjugate momenta of the fields as
\begin{equation}\label{Pi}
\Pi_i (\mathbf{r},t)=\frac{\partial \mathcal{L}}{\partial (\partial_t A_i)}=-\epsilon_0 E_i-P_i=- D_i (\mathbf{r},t),
\end{equation}
\begin{equation}\label{Q}
Q_i=\frac{\partial \mathcal{L}}{\partial (\partial_t X_i)}=\gamma^2(\partial_t X_i+V_0\,\partial_{x}X_i),
\end{equation}
where $P_i (\mathbf{r},t)=\epsilon_0\int_{0}^{\infty} d\nu f_{ij} (\nu,t)\,X^j (\mathbf{r},t,\nu)$ is the polarization component while $ D_i $ is the displacement vector. Enforcing the following commutation relations at equal times,
\begin{equation}\label{QA}
[A_i (\mathbf{r},t),\Pi_j (\mathbf{r}',t)]=i\hbar\,\delta_{ij}\,\delta (\mathbf{r}-\mathbf{r}'),
\end{equation}
\begin{equation}\label{QX}
[X_{i} (\mathbf{r},t,\nu),Q_{j} (\mathbf{r}',t,\nu')]=i\hbar\,\delta_{ij}\,\delta (\mathbf{r}-\mathbf{r}')
\delta (\nu-\nu').
\end{equation}
we apply the process of quantization to the system we are studying.
Now from Euler-Lagrange equations we find the equations of motion for the electromagnetic and matter fields respectively as
\begin{equation}\label{EA}
\frac{1}{c^2}\partial^2_t\mathbf{A}+\nabla\times\nabla\times\mathbf{A} = \mu_0 [\partial_t\mathbf{P}-\nabla\times (\mathbf{v}\times\mathbf{P})],
\end{equation}
and
\begin{eqnarray}\label{EX}
&&\gamma^2(\partial^2_t X_i+2V_0\,\partial_{t}\partial_{x}\,X_i +V_0^2\partial^2_{x}X_i)+\nu^2\,X_i=\nonumber\\ 
&&-\epsilon_0f_{ji}(\nu,t)\,[\partial_t\mathbf{A}-\mathbf{v}\times (\nabla\times\mathbf{A})]_j= -\epsilon_0f_{ji}(\nu,t)\,(\mathbb{D}_t \mathbf{A})_j,\nonumber\\ 
\end{eqnarray}
where $\mathbb{D}_t=\partial_t-\mathbf{v}\times\nabla\times\cdot$, is defined for for the sake of simplicity and convenience in notation. The equation (\ref{EA}) in space-frequency is written as
\begin{equation}\label{FEA}
-\frac{\omega^2}{c^2}\,\mathbf{A}+\nabla\times\nabla\times\mathbf{A} = -i\omega\mu_0 [\mathbf{P}+\frac{1}{i\omega}\nabla\times(\mathbf{v}\times\mathbf{P})]
=-i\omega\mu_0 \tilde{\mathbb{D}}\mathbf{P},
\end{equation}
where $\tilde{\mathbb{D}}=1+\frac{1}{i\omega}\nabla\times (\mathbf{v}\times\cdot)$ and
\begin{equation}\label{X-Expansion}
X_j (\mathbf{r},t,\nu)=\int dk_x  X_j (\mathbf{r}_\bot,k_x,t,\nu) e^{-i k_xx}.
\end{equation}
Inserting Eq.(\ref{X-Expansion}) into Eq.(\ref{EX}) leads to
\begin{widetext}
\begin{eqnarray}\label{EXm}
\gamma^2\partial^2_t X_j(k_x)+2i\gamma^2 k_xV_0\,\partial_{t}X_j(k_x) +(\nu^2-\gamma^2 k_x^2V_0^2) X_j(k_x)=-\epsilon_0f_{ij}(\nu,t)\,(\mathbb{D}_t \mathbf{A})_i(k_x),
\end{eqnarray}
\end{widetext}
with the formal solution
\begin{eqnarray}\label{SX}
X_j(\mathbf{r}_\bot,k_x,t,\nu)&&=X^{N}_{j} (\mathbf{r}_\bot,k_x,t,\nu)-\nonumber\\
 \epsilon_0\int dt' \,G(t-t', && k_x,\nu)  \,f_{ij}(\nu,t')\,(\mathbb{D}_{t'} \mathbf{A})_{i}(k_x),\nonumber\\  
\end{eqnarray}
where the Green's function is given by
\begin{equation}\label{Green-0}
G(t-t';k_x,\nu)=e^{i k_xV_0 (t-t')}\,\frac{\sin[(\frac{\nu}{\gamma}(t-t')]}{\nu}\,\theta(t-t'),
\end{equation}
and the homogeneous solution $X^{N}_{i,m}$, interpreted as a noise field, is
\begin{eqnarray}\label{HOMO}
 X^{N}_{i} (\mathbf{r}_\bot,k_x,t,\nu)=&&a^{\dag}_{i}(\mathbf{r}_\bot,k_x,\nu)e^{i(\nu-k_xV_0)t}\nonumber\\
 &&+a_{i}(\mathbf{r}_\bot,k_x,\nu)e^{-i(\nu+k_xV_0)t}.
\end{eqnarray}
The commutation relation between $a$ and $a^{\dag}$  operators is defined as
\begin{eqnarray}\label{amamdag}
[a_{i}(\mathbf{r}_\bot,k_x,\nu),a^{\dag}_{j}(\mathbf{r}'_\bot,k'_x,\nu')]= &&  \nonumber  \\
 \frac{\hbar}{4\pi\gamma^2\nu}\,\delta_{ij} \,\delta(k_x-k'_x)  \delta(\nu && -\nu') \,\delta(\mathbf{r}_\bot-\mathbf{r}'_\bot)
\end{eqnarray}
The Hamiltonian corresponding to the noise field $X^{N}$ in the body frame is the thermal bath defined by
\begin{eqnarray}
H_B &=& \frac{1}{2}\int_0^\infty d\nu\int d\mathbf{r}\,[(\partial_t \mathbf{X}_\nu)^2+\nu^2 \mathbf{X}^2_\nu]\nonumber \\
   &=& \sum_{j}\int_0^\infty d\nu\int_0^\infty dk_x\int d\mathbf{r}_\bot\,\nu^2 (\hat{a}^\dag_{j}\hat{a}_{j}+\hat{a}_{j}\hat{a}^\dag_{j})\nonumber \\
   &=& \sum_{j}\int_0^\infty d\nu\int_0^\infty dk_x\int d\mathbf{r}_\bot\, \frac{\hbar\nu\gamma^2}{2} (\hat{b}^\dag_{j}\hat{b}_{j}+\hat{b}_{j}\hat{b}^\dag_{j}),\nonumber \\ 
\end{eqnarray}
where in the last equality we have defined the normalized ladder operators $\hat{b}_{j}(\hat{b}^\dag_{j})=\sqrt{\frac{2\nu\gamma^2}{\hbar}}\hat{a}_{j}(\hat{a}^\dag_{j})$.
\begin{eqnarray}
  \langle \hat{b}^\dag_{i}\hat{b}_{j} \rangle &=& tr[\hat{\rho}\, \hat{b}^\dag_{i}\hat{b}_{j}]=\delta_{ij}\delta(\nu-\nu')
  \delta(\mathbf{r}_\bot-\mathbf{r}'_\bot)\frac{1}{e^{\frac{\hbar\nu}{k T}}-1},\nonumber\\
\end{eqnarray}
in which $n_{T}(\omega)=[\exp(\hbar\omega/k T)-1]^{-1}$ is the thermal mean number of photons. From Eq. (\ref{SX}) and the definition of the polarization we have
\begin{eqnarray}\label{PO}
 P_{k} (\mathbf{r}_\bot,k_x,t)&&=P^N_{k} (\mathbf{r}_\bot,k_x,t)-\nonumber \\ 
 \epsilon_0^2\int dt'\int_{0}^{\infty} && d\nu\,f_{ii}^2 (\nu)\,G(t-t';k_x,\nu)\,
(\mathbb{D}_{t'} \mathbf{A})_{i},\nonumber \\    
\end{eqnarray}
where $P^N_{k}$ are the fluctuating or noise polarization components defined by
\begin{equation}\label{N}
P^{N}_{k} (\mathbf{r},t)=\epsilon_0\int_{0}^{\infty}
d\nu\,f_{ki}(\nu)\,X^N_i (\mathbf{r},\nu,t).
\end{equation}
Also, defining response function as
\begin{equation}\label{RES}
\chi^{ee}_{kj}(t-t',k_x) = \epsilon_0\int_{0}^{\infty} d\nu\, G(t-t';k_x,\nu)\,f_{kl} (\nu) f_{jl} (\nu),
\end{equation}
and,
\begin{eqnarray}\label{RES1}
\chi^{ee}_{ii}(t-t',k_x) &=& \epsilon_0\int_{0}^{\infty} d\nu\, G(t-t';k_x,\nu)\,f^{2}_{ii} (\nu),\
\end{eqnarray}
with the following Fourier transforms
\begin{eqnarray}\label{RES2}
\chi^{ee}_{zz}(\omega,m) &=& \epsilon_0\int_{0}^{\infty} d\nu\, \frac{f^{2}_{zz} (\nu)}{\gamma(\frac{\nu^2}{\gamma^2}-(\omega-k_xV_0)^2)},
\end{eqnarray}
\begin{equation}\label{kapa-0}
\chi^{0}_{kk}(\omega) =\epsilon_0\int_{0}^{\infty} d\nu\, \frac{f^{2}_{kk} (\nu)}{\gamma(\frac{\nu^2}{\gamma^2}-\omega^2)},
\end{equation}
one could observe that this recent relation leads to
\begin{equation}\label{kapa-0}
\frac{f^2_{kk} (\gamma\nu)}{\nu}=\frac{2}{\pi\epsilon_0}\,\mbox{Im}[\chi^0_{kk} (\nu)].
\end{equation}
Now using Eq. (\ref{RES2}) one can find the relations between the response functions in body and Lab frames as
\begin{eqnarray}\label{connection}
\chi^{ee}_{ii}(\omega,k_x)=\chi^{0}_{ii}(\omega-k_xV_0)
\end{eqnarray}
Also from Eqs. (\ref{PO},\ref{RES2}) we find
\begin{eqnarray}\label{PO-compact}
&& P_{z} (\mathbf{r},\omega)=P^{N}_{z} (\mathbf{r},\omega)+\epsilon_0\,\chi^{ee}_{zz} (\omega,-i\partial_{x})\,(\mathbb{D}\mathbf{E})_{z},\nonumber\\
&& P_{x} (\mathbf{r},\omega)=P^{N}_{x} (\mathbf{r},\omega)+\epsilon_0\,\chi^{ee}_{xx} (\omega,-i\partial_{x})\,(\mathbb{D}\mathbf{E})_{x},\nonumber\\
&& P_{y} (\mathbf{r},\omega)=P^{N}_{y} (\mathbf{r},\omega)+\epsilon_0\,\chi^{ee}_{yy} (\omega,-i\partial_{x})\,(\mathbb{D}\mathbf{E})_{y},
\end{eqnarray}
or in compact form
\begin{equation}\label{matrix}
\mathbf{P} (\mathbf{r},\omega)=\mathbf{P}^{N} (\mathbf{r},\omega)+\epsilon_0\,\boldsymbol{\chi}^{ee} (\omega,-i\partial_{x})\cdot\mathbb{D}\mathbf{E}.
\end{equation}
In addition, from Eqs. (\ref{PO-compact},\ref{matrix}) one could derive the following relation
\begin{equation}\label{FEA}
\Bigl\{\nabla\times\nabla\times\,-\frac{\omega^2}{c^2}\mathbb{I}-\frac{\omega^2}{c^2}\tilde{\mathbb{D}}\cdot
\boldsymbol{\chi}^{ee}(\omega,-i
\partial_x)\cdot\mathbb{D}\Bigl\}\cdot\mathbf{E}=\mu_0\omega^2\tilde{\mathbb{D}}\mathbf{P}^N.
\end{equation}
In Eq. (\ref{FEA}), the appearance of operators $\mathbb{D}$ and $\tilde{\mathbb{D}}$ indicates the complexity of this relation. However, for small velocities where $v/c \ll 1$, we can make approximate substitutions of $\mathbb{D}$, $\tilde{\mathbb{D}}$, and $\gamma$ with unity, as these terms become negligible in this regime.
\section{Radiation Power}
While both classical and quantum systems undergo energy dissipation during friction, there is a notable distinction in the case of quantum friction as it occurs without physical contact. Unlike classical friction, which involves direct interaction between surfaces, quantum friction is a non-contact phenomenon. Despite this distinction, it is widely acknowledged that quantum friction results in energy dissipation as well. \cite{mandel2016quantum,Kisiel2015}.  Thus, if quantum friction exists in a system, then the radiation power of moving bodies would be crucial in determining its presence. This is because quantum friction is a theoretical concept that arises from the interaction between quantum mechanical particles and their environment, and it is still unclear whether it actually exists in reality. However, if quantum friction does exist, it could potentially manifest as a measurable effect on the radiation emitted by moving bodies, making the study of radiation power a key factor in verifying the existence of this phenomenon.
\par 
Using Eqs. (\ref{N},\ref{HOMO}) one could find that
\begin{eqnarray}
P^{N}_{i} (\mathbf{r}_\bot,k_x,\omega)=&&2\pi\epsilon_0 f_{ii} (k_xV_0-\omega)\,a^{\dag}_{i} (\mathbf{r}_\bot,k_xV_0-\omega)\nonumber \\ 
&&+2\pi\epsilon_0 f_{ii} (\omega-k_xV_0)\,
a_{j,m} (\mathbf{r}_\bot,\omega-k_xV_0).\nonumber \\ 
\end{eqnarray}\label{FFLUC}
Furthermore, if the dielectric is held in temperature $T$, then 
\begin{eqnarray}\label{FLUC-DISS}
&&\langle P^{N}_{i}(\mathbf{r},\omega)P^{N\dag }_{i}(\mathbf{r}',\omega')\rangle =\nonumber \\ 
&&\hbar\epsilon_0\int dk_x e^{i k_x (x-x')}\Gamma_{ii}(\omega,k_x)\,\delta(\omega-\omega')\,\delta(\mathbf{r}_\bot-\mathbf{r}_\bot'),\nonumber \\ 
\end{eqnarray}
where $\Gamma_{ij}$ is defined by
\begin{eqnarray}\label{Gama}
\Gamma_{ii} (\omega,k_x) = 2\mbox{Im}[\chi^{0}_{ii} (k_xV_0-\omega)] a_T (k_xV_0-\omega),
\end{eqnarray}
and $ a_T (\omega)=\coth(\hbar\omega/2k_B T)=2[n_T (\omega)+\frac{1}{2}]$.
\par 
To calculate the radiation power, which is defined as follows,
\begin{equation}\label{Power}
\langle\mathcal{P}\rangle=-\int_{V} d\mathbf{r}\langle(\partial_t\mathbf{P}-\nabla\times (\mathbf{v}\times\mathbf{P}))\cdot (\mathbf{E}+\mathbf{v}\times\mathbf{B})\rangle,
\end{equation}
we note that in the regime of  small velocities $(v/c\ll 1)$, the radiation power could be derived as \cite{PhysRevA.89.032124}
\begin{equation}\label{e38}
\langle\mathcal{P}\rangle=-\int_{V_s} d\mathbf{r}\langle\partial_t\mathbf{P}\cdot\mathbf{E}\rangle.
\end{equation}
in which
\begin{eqnarray}\label{e39}
E_i (\mathbf{r},\omega) &=& E_{0,i} (\mathbf{r},\omega)+\mu_0\omega^2\int d\mathbf{r}'\,G_{0,ii} (\mathbf{r},\mathbf{r}',\omega)\,P^{N}_{i} (\mathbf{r}',\omega),\nonumber\\
P_i (\mathbf{r},\omega) &=& P^N_i (\mathbf{r},\omega)+\epsilon_0 \int dx'\chi^{ee}_{ii} (\omega,-i\partial_{x'}) E^{N}_{i} (x-x',\mathbf{r}_\bot,\omega).\nonumber\\
\end{eqnarray}
Moreover, from \cite{PhysRevA.89.032124}, we realize that the radiated (absorbed) power of the matter field in the presence of electromagnetic vacuum field can be written as,
\begin{equation}\label{power1}
  \langle\mathcal{P}\rangle=\int_{V} d\mathbf{r}\int\int_{-\infty}^{\infty}\frac{d\omega}{2\pi}\frac{d\omega'}{2\pi}e^{-i(\omega+\omega')t}
  (i\omega)\langle \mathbf{P}(\mathbf{r},\omega)\cdot\mathbf{E}(\mathbf{r},\omega')\rangle,
\end{equation}
Therefore, using Eqs.  (\ref{matrix},\ref{FEA}) we find the following expression for the power
\begin{widetext}
\begin{eqnarray}\label{power2}
  \langle\mathcal{P}\rangle =&& \int_{V} d\mathbf{r}\int\int_{-\infty}^{\infty}\frac{d\omega}{2\pi}\frac{d\omega'}{2\pi}e^{-i(\omega+\omega')t}
  (i\omega)\bigg[\mu_0 \omega'^{\,2}\int_{V}d\mathbf{r}'G_{ij}(\mathbf{r},\mathbf{r}',\omega')\langle P_i (\mathbf{r},\omega)
  P_j (\mathbf{r}',\omega')\rangle\nonumber\\
  &&+\epsilon_0\int dx'\chi^{ee}_{ii} (\omega,-i\partial_{x'})
  \langle E^{N}_{i} (x-x',\mathbf{r}_\bot,\omega)E_i (x,\mathbf{r}_\bot',\omega')\rangle|_{\mathbf{r}'\rightarrow\mathbf{r}}\bigg].\nonumber\\
\end{eqnarray}
\end{widetext}
In addition, considering the next relation
\begin{equation}\label{dyad}
\langle E_j (\mathbf{r},\omega)E_i (\mathbf{r}',\omega')\rangle=\frac{2\hbar\omega^2}{\epsilon_0 c^2} \mbox{Im}G_{ji} (\mathbf{r},\mathbf{r}',\omega)
\,\delta(\omega+\omega')\,a_{T_0} (\omega),
\end{equation}
while employing Eq. (\ref{FLUC-DISS}), one could further expand the  radiation power even
\begin{widetext}
\begin{eqnarray}\label{power3}
  \langle\mathcal{P}\rangle &=& \frac{\hbar}{2\pi c^2}\int_{V} d\mathbf{r}\int_{0}^{\infty} d\omega \int dx'\,\omega^3
   \bigg[\int \frac{dk_x}{2 \pi}[2a_{T} (\omega-k_xV_0 )\mbox{Im}\chi^{0}_{ii}(\omega-k_xV_0)\mbox{Im}\,G_{ii}(\mathbf{r},\mathbf{r}',\omega)e^{-ik_x(x-x')}]\nonumber\\
   &&+\int \frac{dk_x}{2\pi}[-a_{T_0}(\omega)\mbox{Im}\chi^{0}_{ii}(\omega-k_xV_0)\mbox{Im}\,G_{ii}(x,x-x',\mathbf{r}_\bot,\mathbf{r}_\bot',\omega)e^{-ik_x(x')}]\bigg ]_{\mathbf{r}'\rightarrow\mathbf{r}}.\nonumber\\
\end{eqnarray}
\end{widetext}
Now, by choosing a proper diadic Green's function $ G_{ij} $, one can model the geometry of the system.  The relevant dielectric function can be written as,
\begin{equation}\label{e}
\varepsilon(\mathbf{r},\omega) = \left\{
\begin{array}{cc}
\varepsilon(\omega) \qquad\quad  z\leqslant 0 \\
1 \qquad\qquad  z> 0,
\end{array}
\right.
\end{equation}
while the diadic Green's function for this specific geometry has been calculated as \cite{PhysRevB.11.1392},
\begin{eqnarray}\label{green}
G_{ii}(\mathbf{r},\mathbf{r}',\omega)&=&\int_{0}^{\infty}\frac{d^2k_\Vert}{(2\pi)^2}e^{i\mathbf{k_\Vert}.
(\mathbf{r}_\Vert-\mathbf{r}_\Vert')}D_{ii}(z,z',\omega)
\end{eqnarray}
where
\begin{eqnarray}\label{green0}
D_{xx}(z,z',\omega)&=&\frac{k_x^2}{k_\Vert ^2}\,g_{xx}(\mathbf{k}_\Vert, \omega\vert z,z')+\frac{k_y^2}{k_\Vert ^2}\,g_{yy}(\mathbf{k}_\Vert, \omega\vert z,z') \nonumber\\
D_{yy}(z,z',\omega)&=&\frac{k_y^2}{k_\Vert ^2}\,g_{xx}(\mathbf{k}_\Vert, \omega\vert z,z')+\frac{k_x^2}{k_\Vert ^2}\,g_{yy}(\mathbf{k}_\Vert, \omega\vert z,z') \nonumber\\ 
D_{zz}(z,z',\omega)&=&g_{zz}(\mathbf{k}_\Vert, \omega\vert z,z')
\end{eqnarray}
and,
\begin{eqnarray}\label{green1}
g_{xx}(\mathbf{k}_\Vert, \omega\vert z,z')&=&-\frac{2\pi i k c^2}{\omega^2}\big [\frac{k_1+\varepsilon(\omega)k}{k_1-\varepsilon(\omega)k}\, e^{ik(z+z')}+e^{ik\vert z-z'\vert} \big ]\nonumber\\
g_{yy}(\mathbf{k}_\Vert, \omega\vert z,z')&=&\frac{2\pi i}{k}\big[\frac{k_1+k}{k_1-k}\,e^{ik(z+z')}-e^{ik\vert z-z'\vert}\big]\\\nonumber
g_{zz}(\mathbf{k}_\Vert, \omega\vert z,z')&=&\frac{2\pi i k_\Vert^2 c^2}{k\omega^2}\big [\frac{k_1+\varepsilon(\omega)k}{k_1-\varepsilon(\omega)k}\, e^{ik(z+z')}-e^{ik\vert z-z'\vert} \big ]\\\nonumber
&&+\frac{4\pi c^2}{\omega^2}\delta(z-z')
\end{eqnarray}
in which $k=(\frac{\omega^2}{c^2}-k_\Vert^2)^{\frac{1}{2}}$ and  $k_1=-(\frac{\varepsilon(\omega)\omega^2}{c^2}-k_\Vert^2)^{\frac{1}{2}}$.
Here, we emphasize that an interesting limiting case is when $\varepsilon(\omega)\rightarrow \infty$ for $z<0$, i.e., the semi-infinite space is an ideal conductor. In this case, we can easily find the imaginary parts of $g_{xx}(\mathbf{k}_\Vert, \omega\vert z,z')$, $g_{yy}(\mathbf{k}_\Vert, \omega\vert z,z')$ and $g_{zz}(\mathbf{k}_\Vert, \omega\vert z,z')$ for $\mathbf{r}\rightarrow\mathbf{r}'$, as
\begin{eqnarray}\label{green2}
\mbox{Im}\,g_{xx}(\mathbf{k}_\Vert, \omega\vert z,z)&=&-\frac{2\pi k c^2}{\omega^2}[1-\cos(2kz)]  \\ \nonumber
\mbox{Im}\,g_{yy}(\mathbf{k}_\Vert, \omega\vert z,z)&=&-\frac{2\pi}{k}[1-\cos(2kz)] \\ \nonumber
\mbox{Im}\,g_{zz}(\mathbf{k}_\Vert, \omega\vert z,z)&=&-\frac{2\pi k_\Vert^2c^2}{k\omega^2}[1+\cos(2kz)]
\end{eqnarray}
\begin{figure*}
\begin{center}
\subfigure[]{\includegraphics[width=8cm]{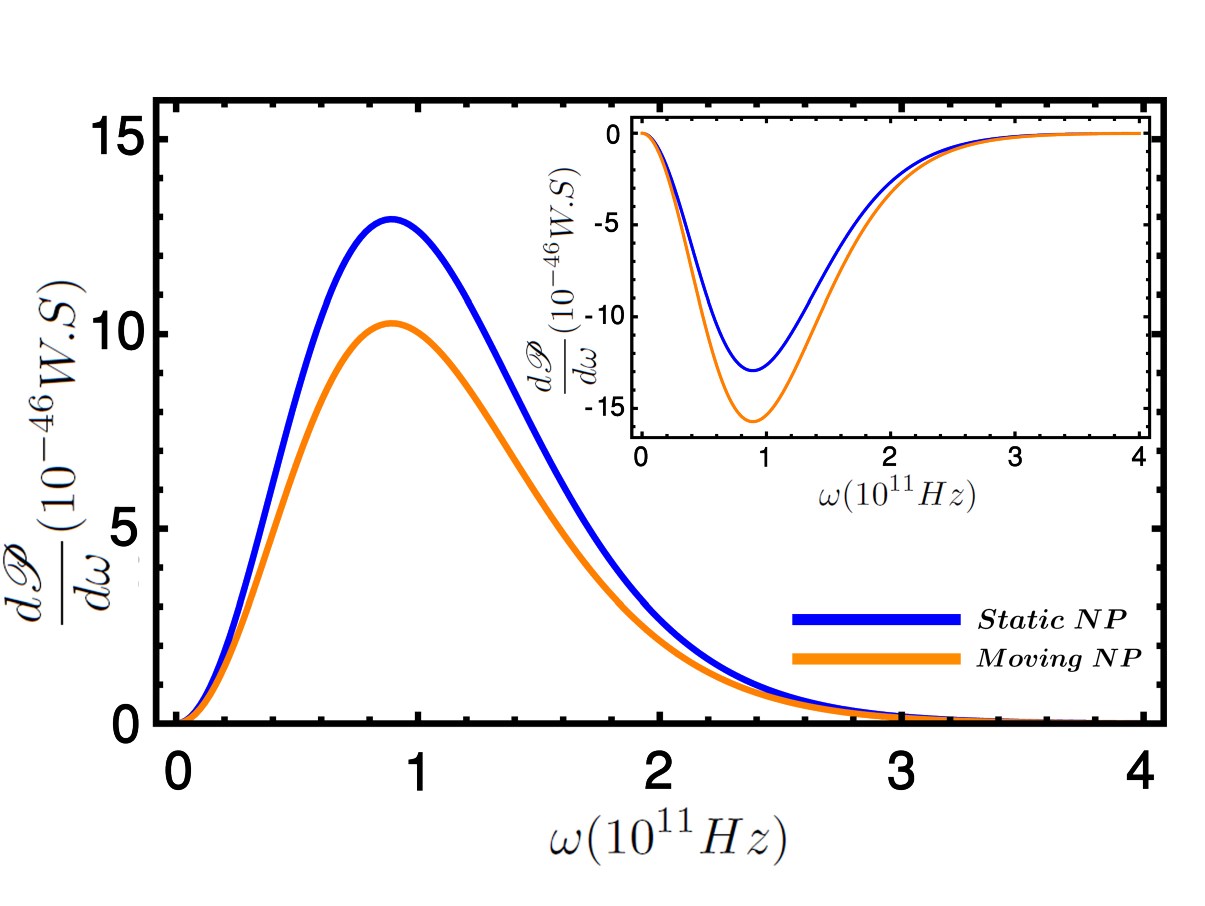}\label{2a}}
\subfigure[]{\includegraphics[width=8cm]{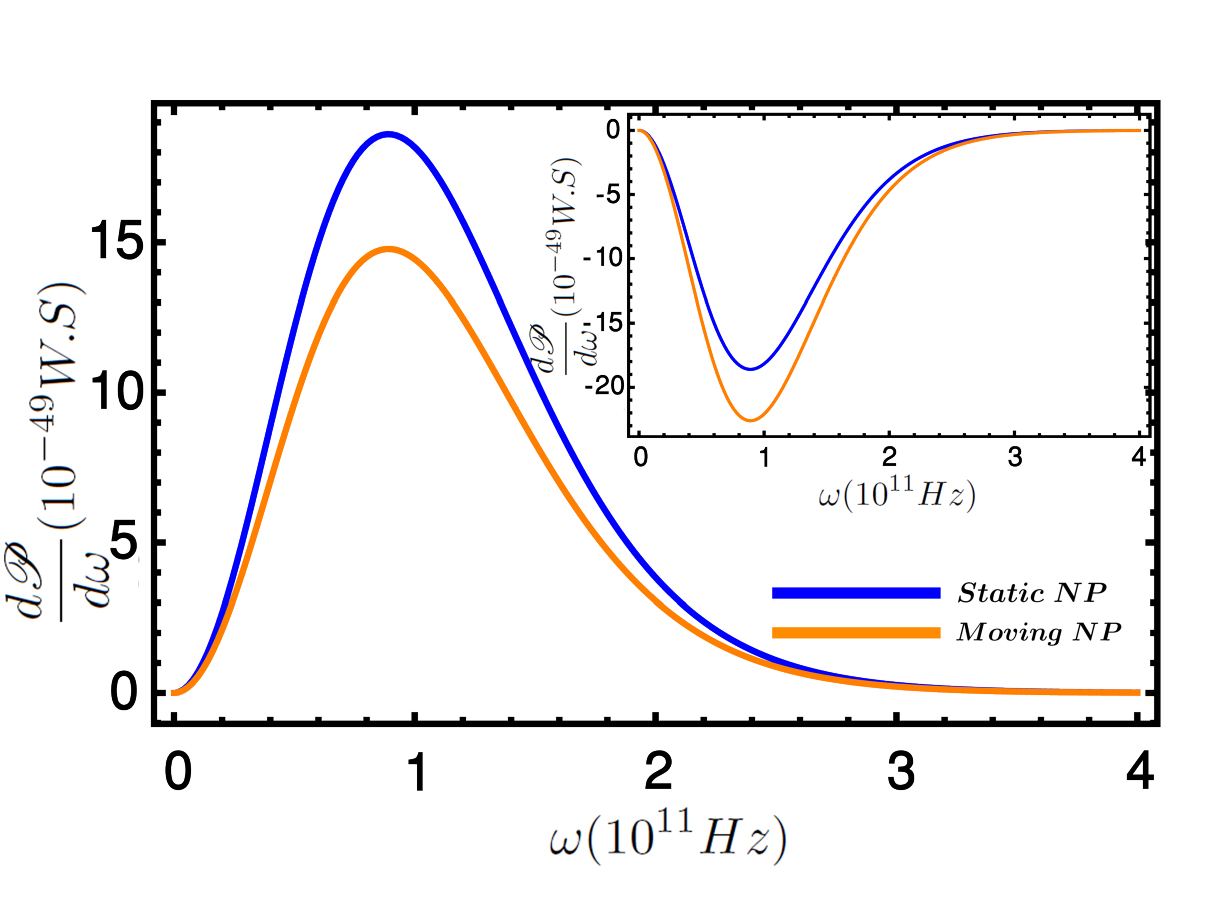}\label{2b}}
\end{center}
\caption{The spectral radiation power for two different nanoparticles, namely (a) Gold and (b) Graphite, with a radius of $a=10 nm$ at a distance of $ z=10^{-4} m $ from the surface and $\sigma_0\approx1.6\times 10^7 \Omega^{-1}m^{-1}, \sigma_0\approx2.3\times 10^4 \Omega^{-1}m^{-1}$ respectively, is depicted. The blue and orange lines represent the spectrum of radiation power for the static and moving particles, respectively, in the presence of a semi-infinite space. In each panel, the temperatures of the particle and its medium are assumed to be 1.11 Kelvin and 1.12 Kelvin, respectively, indicating that the environment is at a higher temperature compared to the particle. A subfigure is also included where the temperatures of the particle and its medium are assumed to be 1.12 Kelvin and 1.11 Kelvin, respectively.
 }\label{fig2}

\end{figure*}
In addition, the dielectric function of metals can be described approximately by Drude model in term of the DC conductivity $\sigma_0$ as
\begin{equation}\label{e10}
\epsilon(\omega)\approx\frac{4i\pi\sigma_0}{\omega},
\end{equation}
Moreover, using the small radius expansion of the Mie scattering coefficient $(a/\lambda\ll 1)$, we find
\begin{equation}\label{e11}
\alpha(\omega)\approx a^3\frac{\epsilon(\omega)-1}{\epsilon(\omega)+2},
\end{equation}

from which one could finally derive the radiation power of the system as
\begin{widetext}
\begin{eqnarray}\label{power4}
\langle\mathcal{P}\rangle =\frac{\hbar}{2\pi c^2}\int_0^\infty\frac{d^2k_\Vert}{(2\pi)^2}\int_{0}^{\infty} d\omega\,\omega^3\bigg[2\mbox{Im}\alpha(\omega-k_xV_0)[a_{T} (\omega-k_xV_0)-a_{T_0}(\omega)]\nonumber\\
   \big [\mbox{Im}\,D_{xx}(z,z',\omega)+\mbox{Im}\,D_{yy}(z,z',\omega)+\mbox{Im}\,D_{zz}(z,z',\omega)\big] \bigg].
\end{eqnarray}
\end{widetext}
\begin{figure*}
\begin{center}
\subfigure[]{\includegraphics[width=8cm]{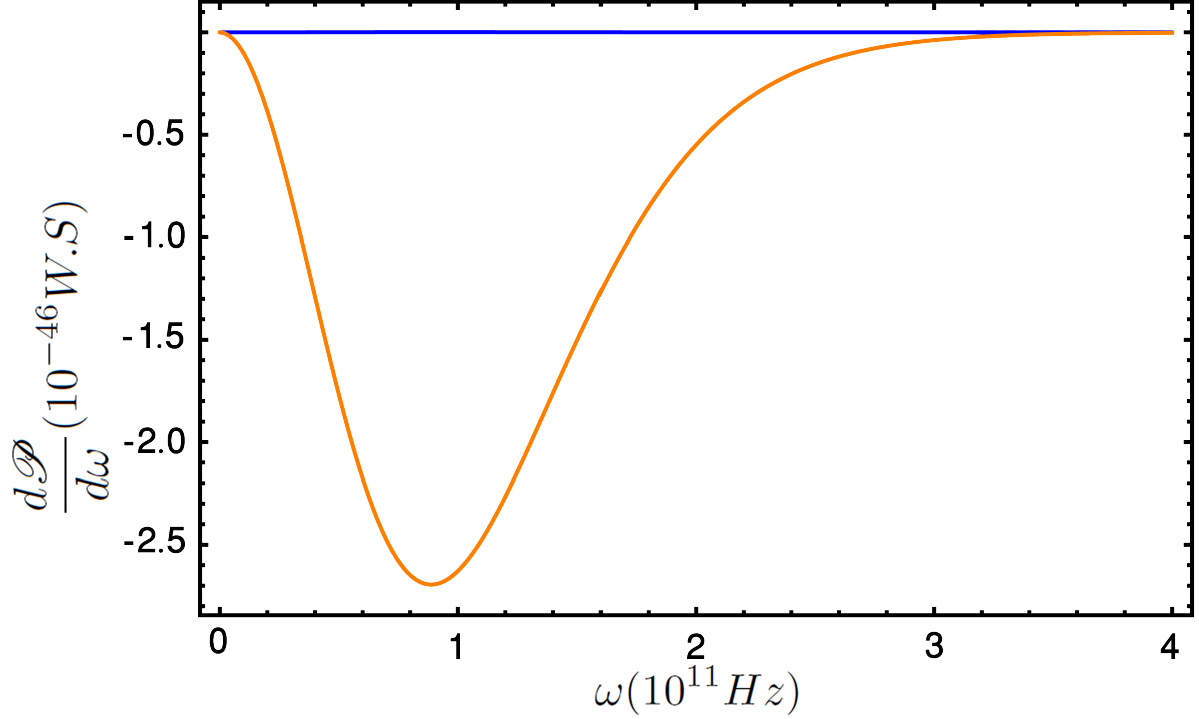}\label{3f}}
\subfigure[]{\includegraphics[width=8cm]{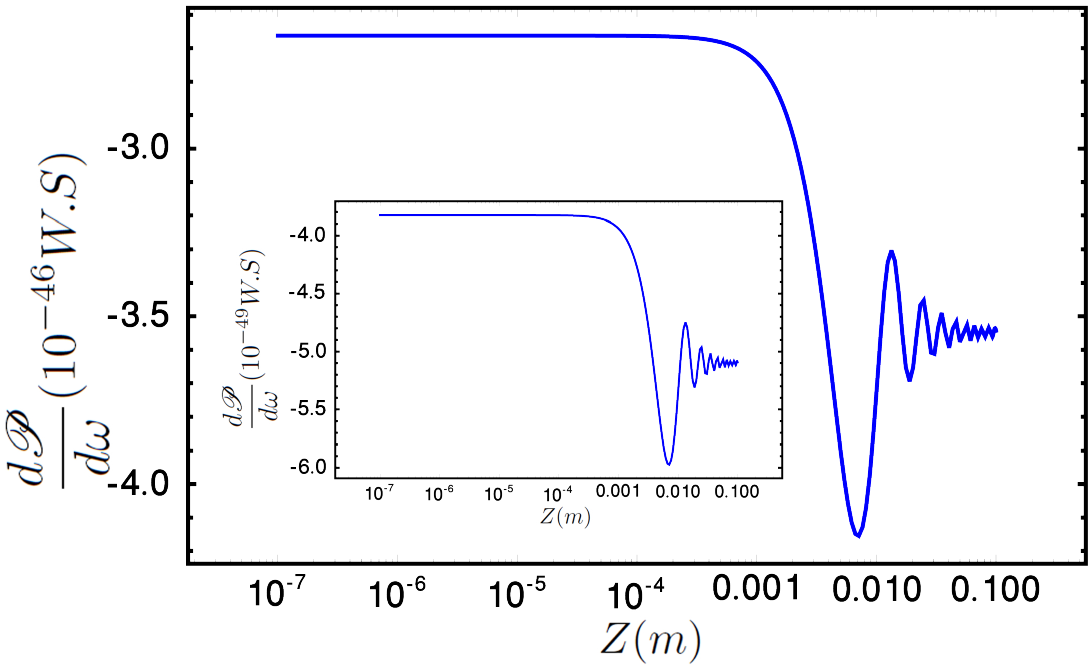}\label{NF}}
\end{center}
 \caption{In panel (a), the spectral radiation power for gold nanoparticles with a radius of $a=10nm$ and a conductivity of$(\sigma_0\approx1.6\times 10^7 \Omega^{-1}m^{-1})$ is analyzed. The blue and orange lines represent the spectral radiation power for a static and a moving particle, respectively. In panel (b), the analysis focuses on the peak point of the spectrum at $ \omega=0.9*10^{11} $. The impact of the distance to the surface is investigated for gold NPs (and graphite NPs in a subfigure). In both panels, the nanoparticles are assumed to be in thermal equilibrium at a temperature of $T=1.11 K$ with their surrounding medium.}
\end{figure*}

\par

It is noteworthy to analyse Eq. (\ref{power4}) in certain special cases, such as in light of Einstein's theory of special relativity, where physics remains invariant for any object moving at a constant speed in any inertial frame. For instance, when considering a single nanoparticle (NP) in a vacuum, the radiation power derived in Eq. (\ref{power4}) is the same for both moving and stationary NPs, which is not surprising. However, when a semi-infinite dielectric bulk is present near a moving NP, the situation becomes different. 
\par
In Fig.(\ref{fig2}), we present a comparison of two distinct scenarios. In panel 2a, the nanoparticle temperature is elevated by 0.01 K relative to the environment, while in panel 2b, the opposite condition is observed. For the stationary NP, both figures display identical values but with opposite signs. However, in the case of the moving NP, panel 2a exhibits a higher peak, whereas panel 2b indicates a lower peak. The total radiation power in both instances encompasses a negative component stemming from quantum friction dissipation. As anticipated, a lower peak is observed when the thermal radiation power is positive, and conversely, a higher peak emerges when the thermal radiation power is negative.
\par
This negative term can be attributed to the radiation power resulting from the dissipation of non-contact friction forces. This term has been isolated and illustrated in Fig.(\ref{3f}) under the assumption of equivalent temperatures for both the NP and its surrounding environment. Fig.(\ref{NF}) showcases the peak frequency of the spectral radiation power for a moving NP in thermal equilibrium with its environment relative to the distance between the NP and the semi-infinite bulk. Evidently, two distinct regimes are discernible in the spectrum, corresponding to near-field and far-field conditions.

\section{Conclusion}
By employing the technique of field quantization, we have derived the radiation power of a moving NP alongside a semi-infinite bulk in Eq. (\ref{power4}). Calculating the dissipation caused by quantum friction of the NP is a straightforward process using this equation. The total radiation power in Eq. (\ref{power4}) of a moving NP consists of the radiations resulting from the NP's temperature and the dissipation caused by the friction force. The thermal radiation power can be positive or negative depending on the NP and the environment's temperature, but the dissipation caused by quantum friction force is always negative. By comparing the radiation power of a stationary NP with that of a moving one, it is easy to determine the dissipation power.

\par
The negative term, which has be derived from the total radiation power, is a direct consequence of the dissipation of the non-contact friction force. In essence, when a moving NP is subjected to motion, it experiences a non-contact friction force that leads to energy dissipation. This energy dissipation can be identified as a negative term in the total radiation power. This radiation power, resulting from the dissipation of non-contact friction force, is a distinctive characteristic of the quantum mechanical depiction of friction.

\section*{Acknowledgments} 
The authors are grateful to University of Hormozgan, University of Kashan and Shahrood University of Technolog for the support of this work. Special Thanks to professor Fardin Kheirandish for his valuable comments.

\nocite{*}

\bibliography{apssamp}

\end{document}